\documentclass[prd,twocolumn,showpacs,superscriptaddress,nofootinbib]{revtex4}
\usepackage{amssymb,amsmath,amsfonts,epsfig, subfigure,graphicx}
\usepackage{epic}

\newcommand{\beqs}{\begin{equation*}}
\newcommand{\beq}{\begin{equation}}
\newcommand{\eeqs}{\end{equation*}}
\newcommand{\eeq}{\end{equation}}
\newcommand{\beqas}{\begin{eqnarray*}}
\newcommand{\beqa}{\begin{eqnarray}}
\newcommand{\eeqas}{\end{eqnarray*}}
\newcommand{\eeqa}{\end{eqnarray}}

\newcommand{\eq}[2]{\begin{equation} #1 \label{#2} \end{equation}}

\newcommand{\ka}{\kappa}
\newcommand{\la}{\lambda}
\newcommand{\si}{\sigma}

\newcommand{\La}{\Lambda}

\newcommand{\blist}{\begin{itemize}}
\newcommand{\elist}{\end{itemize}}

\providecommand{\href}[2]{#2}

\DeclareFontFamily{OT1}{rsfs}{}
\DeclareFontShape{OT1}{rsfs}{m}{n}{ <-7> rsfs5 <7-10> rsfs7 <10->rsfs10}{} 
\DeclareMathAlphabet{\mycal}{OT1}{rsfs}{m}{n}

\DeclareMathOperator{\extdm}{d}
\newcommand{\extd}{\extdm \!}

\begin{document}
\title{Lower bound on the spectral dimension near a black hole}
\author{S.~Carlip}
\affiliation{Department of Physics, University of California, Davis, CA 95616, USA}
\email{carlip@physics.ucdavis.edu}
\author{D.~Grumiller}
\affiliation{Institute for Theoretical Physics, Vienna University of Technology, 
Wiedner Hauptstr.~8-10/136, A-1040 Vienna, Austria, Europe}
\email{grumil@hep.itp.tuwien.ac.at}
\date{\today}
 
\begin{abstract}
We consider an evaporating Schwarzschild black hole in a framework in
which the spectral dimension of spacetime varies continuously from four
at large distances to a number smaller than three at small distances, as 
suggested by various approaches to quantum gravity.  We demonstrate 
that the evaporation stops when the horizon radius reaches a scale at 
which spacetime becomes effectively 3-dimensional, and argue that 
an observer remaining outside the horizon cannot probe the properties
of the black hole at smaller scales.  
This result is universal in the sense that it does not depend on the 
details of the effective dimension as a function of the diffusion time.  
Observers falling into the black hole can resolve smaller scales, as can
external observers in the presence of a cosmological constant.  Even in
these cases, though, we obtain an absolute bound $D\geq2$ on the 
effective dimension that can be seen in any such attempt to measure
the properties of the black hole.
\end{abstract}
\maketitle

\section{Introduction}

General relativity describes spacetime as a smooth $d$-dimensional 
manifold.  But while this picture has proven remarkably successful, it 
is quite plausible that it will break down at very small scales.  A quantum 
theory of gravity must be, in some sense, a theory of the quantization 
of spacetime, and there is no reason to expect that a smooth classical
description will hold to all scales.  A central task of quantum gravity
is to investigate alternative small scale descriptions.

Even without a smooth manifold structure, it is often possible
to define an effective dimension of spacetime.  The spectral dimension 
\cite{Ambjorn:2005db,Hughes}, for instance, is determined by the
rate of a diffusion process, and exists for any space on which a random 
walk can be defined.  Such a definition seems tailor-made for thermodynamic
applications, such as the process of black hole evaporation considered 
in the present paper.

Although it is risky to make too strong a claim without a definitive 
quantum theory of gravity, evidence from a number of different approaches 
suggests that the spectral dimension and similar generalized dimensions 
flow from four at large distance scales to two near the Planck scale 
\cite{Carlip:2009km,Carlipx}.  In particular, Causal Dynamical 
Triangulations (CDT) --- a Lorentzian lattice approach to the gravitational 
path integral --- yields a spectral dimension of spacetime, determined 
numerically, of the form \cite{Ambjorn:2005db}
\eq{
\textrm{CDT:}\qquad D_{\rm IR} = 4.0 \pm 0.1
\qquad D_{\rm UV} = 1.80 \pm 0.25\,.
}{eq:1}
Here $D_{\rm IR}$ is the spectral dimension in the limit of infinite
diffusion time, corresponding to an effective dimension at very large
distances, while $D_{\rm UV}$ is the spectral dimension for short
diffusion times, giving an effective dimension at very small distances.   

It is evident that the spacetime dimension at large scales is compatible 
with four, but the dimension at small scales is smaller than four at  
greater than $5\sigma$ significance.  Moreover, \eqref{eq:1} is 
consistent with the suggestion that the small scale structure is 
effectively 2-dimensional.  A reasonably good fit, valid for arbitrary 
values of the diffusion time $\si$, is \cite{Ambjorn:2005db}
\eq{
D_{\rm spec}(\si) = a - \frac{b}{c+\si}
}{eq:5}
where $D_{\rm IR} = a$ and $D_{\rm UV} = a - b/c$.
The constants $b$ and $c$ can be rescaled arbitrarily through changes of 
the lattice spacing, but their ratio and the constant $a$ are both universal.

On the other hand, a recent study using an alternative lattice approach
known as Euclidean Dynamical Triangulations (EDT) leads to a rather 
different result for the small scale dimension \cite{Laiho:2011ya}:
\eq{
\textrm{EDT:}\qquad D_{\rm IR} = 4.0 \pm 0.3
\qquad D_{\rm UV} = 1.46 \pm 0.06
}{eq:2}
In fact, the result \eqref{eq:2} may suggest $D_{\rm UV}=3/2$, a result   
that is also compatible with \eqref{eq:1}.  Amusingly, this is precisely 
the value for which the Bekenstein--Hawking entropy of a $d$-dimensional 
Schwarzschild black hole,
\eq{
S_{\rm BH} \sim E^{(d-2)/(d-3)}\,,
}{eq:3}
coincides with the entropy of a $d$-dimensional CFT,
\eq{
S_{\rm CFT}\sim E^{(d-1)/d}\,.
}{eq:4}

It is thus of interest to see what an evaporating Schwarzschild black hole 
has to say about this issue.  There are various physical scenarios of interest.
For instance, an observer could intend to probe microscopic distances with 
some high-energy scattering experiment.  If the energy deposited in such an 
experiment gets concentrated in a sufficiently small region, then a black 
hole is created.  However, we shall describe a different situation, in which 
the black hole exists already before the experiment is performed, so that 
we do not have to deal with the rather complicated process of black hole 
formation.  We locate an observer outside a black hole, which for simplicity 
we assume to be spherically symmetric.  She then performs some experiment 
permitting her to probe the scale of spherical shells concentrically surrounding 
the black hole.  Heuristically, the black hole horizon hides the interior --- and 
the related short-distance physics --- from an outside observer.   But as the 
black hole evaporates, its horizon shrinks, allowing the observer to probe 
smaller and smaller distances,\footnote{By ``distance'' in this paper we 
always mean the radius of a $(d-2)$-sphere enveloping the center of the 
$d$-dimensional black hole.} thereby gleaning some information about 
the effective dimension at small scales.  An observer desperate for 
information about short-distance physics might even throw herself into 
the black hole to resolve the effective dimension at even smaller scales. 
The question we shall address is whether the dynamics puts any limit on 
either of these processes.

\section{Dilaton black hole}

We use 2-dimensional dilaton gravity to describe 
the (Euclidean) $d$-dimensional Schwarzschild black hole in 
various dimensions (see, e.g., \cite{Grumiller:2002nm}).  This 
description has several advantages: it is simple; it captures the 
full classical and thermodynamical content of the theory 
\cite{Grumiller:2007ju}; and it allows a straightforward analytic 
continuation to arbitrary (even fractal or negative) dimensions.

The dilaton gravity action is given by
\eq{
I = -\frac{1}{2 G_2}\,\int\!\extd^2x\sqrt{g}\,
   \big[XR - U(X)(\partial X)^2 - 2V(X)\big] + I_{\rm b}\,,
}{eq:6}
with gravitational coupling constant $G_2$ and a known boundary 
action $I_{\rm b}$ that is irrelevant to
the current discussion.  The dilaton has a higher-dimensional
interpretation as the surface area; that is, $X(t,r)$ is the area
of the $(d-2)$-sphere at fixed $t$ and $r$ (the orbit of the Killing 
vectors responsible for spherical symmetry).  

The three terms in the bracket also have straightforward higher-%
dimensional meanings.  Each represents a contribution to the 
$d$-dimensional Ricci scalar.  The first term describes the 
intrinsic curvature of the 2-dimensional spacetime.  The third 
term describes the intrinsic curvature of the $(d-2)$-sphere.  The 
second term gives the contribution to curvature arising from
the change of the area of the $(d-2)$-sphere as a function of time 
and radius.  This is the term we shall modify by hand to accommodate 
an effective dimension that changes with the distance from the black 
hole.

For a $d$-dimensional Schwarzschild black hole, the potential 
$U(X)$ is given by
\eq{
U(X) = -\frac{1}{X}\,\frac{d-3}{d-2}\,.
}{eq:7}
It is useful to define functions
\begin{align}
Q(X)&:=Q_0 + \int^X\!\!\!\extd X^\prime\, U(X^\prime)\,, \\ 
w(X)&:=w_0-2\int^X\!\!\!\extd X^\prime\, e^{Q(X^\prime)}V(X^\prime)\,, 
\label{eq:12}
\end{align}
with some arbitrary integration constants $Q_0$ and $w_0$.
The potential $V(X)$ can be obtained from the requirement that 
the model \eqref{eq:6} have a flat ground state \cite{Grumiller:2002nm}:
\eq{
V(X) \propto e^{-2Q(X)}\,U(X)
}{eq:10}
This requirement then yields
\eq{
V(X) \propto X^{(d-4)/(d-2)}\,,
}{eq:21}
with a proportionality constant that sets the physical length scale.

The classical solutions of the field equations  coming from the action
\eqref{eq:6} are then given by
\begin{alignat}{3}
 X &= X(r) &&\mathrm{with}\;  \partial_r X = e^{-Q(X)} \label{eq:lalapetz} \\
 \extd s^2 &=  \xi(r) \,\extd\tau^2 &&+ \frac{\extd r^2}{\xi(r)}
\label{eq:angelinajolie}\\
 &&&\mathrm{with}\;   \xi(X) 
= w(X)e^{Q(X)}\Big(1-\frac{4M}{w(X)}\Big) \,.
\nonumber 
\end{alignat}
They are parametrized by a single constant of motion, the black hole mass $M$.  
The flat ground state property \eqref{eq:10} implies $e^{Q(X)}w(X)=\rm const.$, 
so that the Killing norm $\xi$ is constant for vanishing black hole mass.

\section{Varying effective dimension and black hole evaporation}\label{sec:3}

For fixed dimension $d$, the solution \eqref{eq:lalapetz} with the potential 
\eqref{eq:7} gives a dilaton
\eq{
X \sim r^{d-2}\,.
}{eq:dill}
As we show in appendix \ref{app:B}, the spectral dimension determined 
from the corresponding dimensionally reduced d'Alembertian is $d$.  
This holds even if $d$ is not an integer.   A varying spectral dimension 
might thus reasonably correspond to a varying $d$ in \eqref{eq:7}.

Let us suppose that over some relevant range of scales, the effective 
dimension is a strictly monotonic function of the diffusion time $\si$.
This behavior occurs in both the CDT and EDT simulations described
in the introduction; the CDT fit \eqref{eq:5} is an example of such
a functional dependence.   We now make our two key working
assumptions:
\begin{enumerate}
\item The diffusion time $\sigma$ --- specifically, the diffusion
   time necessary to capture information about the transverse
   space of constant $r$ and $t$ --- is itself a strictly
   monotonic function of the dilaton $X$.
\item The potential \eqref{eq:7} remains valid even when the
dimension depends on the scale; that is,
\eq{
U(X) = -\frac{1}{X}\,\frac{D(X)-3}{D(X)-2}\,.
}{eq:9}
\end{enumerate} 
The first assumption is motivated by the observation that $\si$ 
and $X$ both determine the scale: small diffusion times and 
small values of the dilaton field both correspond to small distances,
while large $\si$ and $X$ both correspond to large distances.  The
second comes from the interpretation of the second term in the action 
\eqref{eq:6}, as described in the paragraph below that equation,
and from the results of appendix \ref{app:B}.  We check this assumption for a
particular potential in appendix \ref{app:A}, and show that the $D(X)$ in
\eqref{eq:9} is in good agreement with the spectral dimension over the whole range of $X$, with a 
maximal deviation of 15\% (see Fig.~\ref{fig:1} in appendix \ref{app:A}).

For the sake of concreteness, we shall assume that the fit 
\eqref{eq:5} is valid and that the diffusion time is a monotonic 
increasing function $f$ 
of the dilaton $X$:
\eq{
D(X) = a - \frac{b}{c+f(X)}
}{eq:8}
In appendix \ref{app:A}, we provide a simple toy model with linear $f$, which 
allows us to elucidate certain aspects of the black hole evaporation.
Our main conclusions are independent of these specific choices,
however; all that matters is that $D(X)$ is some strictly monotonic 
function of $X$ that goes to four for $X\to\infty$ and to some value 
smaller than three for $X\to 0$.

Again we determine $V(X)$ from the flat ground state requirement 
\eqref{eq:10}.  This choice ensures that the theory allows 2-dimensional flat spacetime as solution. 
These choices above should 
be considered as working assumptions; others are conceivable.  However, 
we believe these assumptions are sufficiently well-motivated to warrant 
a study of their consequences.

We make now our key observation.  The function $V(X)$ vanishes if 
$U(X)$ vanishes, which happens precisely for $D(X)=3$, regardless 
of the detailed properties of the effective dimension as a function of 
the dilaton.  The vanishing of $V$ implies, in turn, that $w'$ is 
zero.  Since surface gravity $\ka$ is given by 
\cite{Gegenberg:1994pv,Grumiller:2007ju}
\eq{
\ka = \frac12\, w'\Big|_{X=X_h} \propto V(X_h)
}{eq:11}
where $X_h$ 
is the value of the dilaton evaluated at the horizon, it follows that the 
black hole is extremal if $V(X_h)=0$.   Consequently, the black hole stops
evaporating once the horizon drops to a scale for which $D(X_h)=3$.

As a byproduct, we also learn that specific heat must turn positive 
before the black hole horizon drops to the critical size at which
$D(X_h)=3$.  To see this, note first that the temperature increases
monotonically as long as the effective dimension $D(X)$ is sufficiently 
close to four.  In that region we recover the standard result that 
Schwarzschild black holes have negative specific heat.  Since the
temperature drops to zero smoothly as $D(X_h)\to 3$, it must have 
a maximum at some value $X_h^c$, with $3<D(X_h^c)<4$.  In the 
region $X_h<X_h^c$, the specific heat is therefore positive.  Such a 
behavior might be expected on general grounds for quantum-corrected 
Schwarzschild black holes.

A further byproduct is that a curvature singularity necessarily appears
behind the horizon.  This can be shown as follows.  The 2-dimensional 
Ricci scalar is \cite{Grumiller:2002nm}
\begin{multline}
R = 4M e^{-Q(X)} U^\prime(X) \\
\propto \frac{(D(X)-2)(D(X)-3)-XD^\prime(X)}{X^2(D(X)-2)^2}\,. \label{eq:13}
\end{multline}
The curvature diverges at $X=0$ and $D(X)=2$.  Both loci are always 
within the black hole region, according to the results above.  While the 
existence of a curvature singularity might have been anticipated on 
general grounds from singularity theorems, it is not clear that they 
apply to a situation in which the effective dimension varies.

Interestingly, the result \eqref{eq:13}  implies that not even an 
observer falling into a black hole is able to resolve scales of the 
surface area corresponding to an effective dimension smaller than 
two.  Thus, even if the effective dimension near $X=0$ were given 
by, say, $D_{\rm UV} = 3/2$, no observer would encounter this value
before reaching the singularity.\footnote{Recall that $D(X)$
is essentially a spectral dimension measured with a diffusion
time determined by the area of the space of constant $r$ and
$t$.  One could imagine a \emph{local} observation made by
a freely falling observer at a much smaller scale (with the caveat
that such a process might create another black hole).  
As mentioned at the end of the introduction, such
an observation would not be easily described in the framework
of spherically symmetric dimensional reduction, and our
arguments do not say anything about the possible outcome.
But such a measurement would also not capture the properties
of the black hole, which are our main interest here.}  Instead,
we establish the result that no observer, inside or outside the
horizon, can see a value $D(X)<2$.

\section{Adding a cosmological constant}

In the derivation above, the assumption \eqref{eq:10} of a flat
ground state was crucial.  Let us relax this assumption to allow 
for de Sitter or anti-de Sitter ground states.  This is of interest in
part because our present Universe appears to have a positive 
cosmological constant \cite{Riess:1998cb,Perlmutter:1998np}.   On 
more theoretical grounds, 3-dimensional general relativity has no 
black hole solutions unless a negative cosmological constant is present 
\cite{Banados:1992wn,Schleich}, and one might worry that the critical 
dimension of three derived above may merely reflect this fact.

Given a dilaton gravity model \eqref{eq:6} obtained from the
dimensional reduction of $D$-dimensional Einstein gravity, we can 
add a cosmological constant by a simple shift of the potential $V(X)$ 
\cite{Grumiller:2002nm,Grumiller:2007ju},
\eq{
V_\La (X) = -\la^2 e^{-2Q(X)}\,U(X) + \La N(X) X\,.
}{eq:22}
Here $\la$ is a dimensionful constant that sets the physical scale, 
$\La$ is the cosmological constant, and $N(X)>0$ is a $D$-dependent 
normalization.  Some standard choices are $N(X)=D(X)(D(X)-1)$ and 
$N(X)=1$.  Our conclusions below are independent of the precise 
choice of $N(X)$, as long as it contains no zeros or singularities for 
$D\geq 2$.  We assume that this is the case.
With this change, Hawking evaporation no longer stops at $D(X)=3$, 
but continues until some other value of $D$, which is determined 
by the condition $V_\La(X)=0$, i.e.,
\eq{
\frac{D(X)-3}{D(X)-2} = \frac{\La}{\la^2}\,X^2 N(X) e^{2Q(X)}\,.
}{eq:23}
The key observation is that the right hand side of \eqref{eq:23} is 
always positive (negative) for a positive (negative) cosmological constant.  
This means that \eqref{eq:23} must have a solution with $D(X)>3$ 
for $\La>0$, and with $D(X)<3$ for $\La<0$.

Moreover, it is clear that for $\La<0$ a solution must exist with 
$D(X)>2$.  Indeed, for any finite negative value of $\La$ the right-%
hand side is bounded from below, while the left-hand side is unbounded 
from below as $D\to 2$ from above.  Hence if one plots the two sides
as functions of $X$, the curves must intersect at some $X=X_c$
such that $D(X_c)$ lies between two and three.

Thus, for $\La>0$, Hawking evaporation stops at 
some critical value $X^c$ with $D(X^c)>3$, while for $\La<0$, the 
evaporation stops at a critical value with $2<D(X^c)<3$.
Clearly, if $\Lambda$ is tiny, the critical dimension at which 
evaporation stops is very close to three.  If $\Lambda$ is large and 
negative, on the other hand, an external observer can resolve 
distance scales small enough to correspond to a dimension smaller 
than three.  It remains true, however, that $D=2$ is an absolute
bound for any observer, inside or outside the horizon, regardless of 
the value of the cosmological constant.  This bound is insensitive to 
the details of any of our choices --- the relationship between spectral 
dimension and diffusion time, the relationship between the diffusion 
time and the dilaton field, the normalization of the cosmological 
constant --- as long as these respect the plausible monotonicity 
properties we introduced above.

We conclude that no observer --- outside or inside the black hole
 --- is capable of resolving (radial) distances that correspond to an effective 
dimension smaller than two.  These results provide independent 
evidence in favor of the proposal \cite{Carlip:2009km} that quantum 
gravity should be effectively 2-dimensional at small distance scales.

\acknowledgments 

SC is supported in part by U.S.\ Department of Energy grant DE-FG02-91ER40674.
DG is supported by the START project Y435-N16 of the Austrian Science Fund (FWF). 

\appendix

\section{Dimensional reduction and the spectral dimension}\label{app:B}

Consider any space in which a diffusion process can be defined.  Such a
process is characterized by a heat kernel $K({\bf x}', {\bf x}; \si)$.
The spectral dimension is the dimension measured by the rate of 
diffusion \cite{Ambjorn:2005db,Hughes},
\eq{
D_{\rm spec}(\si) = -2\si\,\frac{\extd}{\extd\si}\,\ln K({\bf x}, {\bf x}; \si)\,.
}{eq:ap2a}
For a flat $d$-dimensional space, the heat kernel is \cite{Vassilevich:2003xt}
\eq{
K({\bf x}', {\bf x}; \si) 
   = (4\pi\si)^{-d/2}\exp\left\{ -\frac{|{\bf x}'-{\bf x}|^2}{4\si}\right\}
}{eq:ap2aa}
and it is easily checked that $D_{\rm spec} = d$.

In the dimensional reduction we have considered here, a $d$-dimensional 
spacetime is treated as if it had only two dimensions.  If the lower 
dimensional model reflects the true physics, though, it must somehow 
capture the full spectral dimension.  To see how this works, consider 
first the case of a flat spacetime, with a dimensionally reduced (Euclidean) 
d'Alembertian
\eq{
\Delta = \partial_\tau^2 + \partial_r^2 
   + (\partial_r \ln{X}) \partial_r + \frac{1}{r^2}{\tilde\Delta}_{d-2} 
}{eq:ap2b}
where the dilaton is the surface area, $X\sim r^{d-2}$, and 
${\tilde\Delta}_{d-2}$ is the Laplacian on the $(d-2)$-sphere. 
The first two terms yield the intrinsic 2-dimensional d'Alembertian, 
while the last two terms
capture information about the remaining $d-2$ dimensions.  We can 
similarly split the dimension $d$ into two parts, one corresponding 
to the 2-dimensional d'Alembertian and one coming from the 
dimensional dependence of the dilaton field on $r$:
\eq{
d = 2 + \frac{\extd\, ({\ln X})}{\extd\, ({\ln r})}  
}{eq:ap2i}
We shall now show that as long as the diffusion time $\sigma$ is not 
too small, this is a good approximation of the spectral dimension
$D_{\rm spec}$, even if $d$ is not an integer.

We are interested in a spherical reduction, in which only the zero 
angular momentum modes are present.  The eigenvalues
of ${\tilde\Delta}_{d-2}$  are $\ell(\ell+d-3)$, so for
these $\ell=0$ modes, the last term in \eqref{eq:ap2b} drops out.  
The operator \eqref{eq:ap2b} is hermitian with respect to the 
integration measure
\eq{
\extd\mu = \extd\tau \extd r\, r^{d-2}
}{eq:ap2c}
and has nonsingular orthonormal eigenfunctions 
\eq{
f_{\omega k}(t,r) = \sqrt{\frac{k}{2\pi}} \,
      r^{\frac{3-d}{2}}e^{i\omega t}J_\nu(kr)  
     \quad\hbox{with\ \ $\nu = \frac{d-3}{2}$}  
}{eq:ap2d}
with eigenvalues $-\omega^2-k^2$.
We can use these to evaluate the heat kernel for $\Delta$ (again with $\ell=0$):
\begin{multline}
\!\!\!\!K_0(r',t',r,t;\si) = \int\!\extd\omega\extd k e^{-\si(k^2+\omega^2)}
                            f^*_{\omega k}(t',r')f_{\omega k}(t,r) \\
  = \frac{1}{\sqrt{2\pi \si}}(rr')^{\frac{3-d}{2}}e^{-(t'-t)^2/4\si}
  \frac{1}{2\si}e^{-({r'}^2+r^2)/4\si} I_\nu\left(\frac{r'r}{2\si}\right)
\label{eq:ap2e}
\end{multline}
It may be checked that this is equivalent to the angular average of 
the flat space heat kernel \eqref{eq:ap2aa}
over a $(d-2)$-sphere of fixed $r$ and $\tau$.

To obtain a spectral dimension, we need a logarithmic derivative of
this quantity.  
For small $\si$, the argument of the modified Bessel 
function is large, and we can use the asymptotic behavior
$
I_\nu(z)\sim 
e^z/\sqrt{2\pi z} 
$.
It is then easy to check that $D_{\rm spec}\sim 2$: in this limit,
the heat kernel does not see the higher dimensional space.
This means that the diffusion time must not be too small (as 
compared to $X^{2/(d-2)}$), since otherwise the s-waves are
not sensitive to the higher dimensions, but merely feel the 
presence of the time and radial coordinates.  This observation provides 
an independent motivation for our first working assumption in 
section \ref{sec:3}.
For large $\si$, on the other hand, we can exploit the asymptotic
behavior of the modified Bessel function at small argument,
\eq{
I_\nu(z)\sim \frac{(\frac{1}{2}z)^\nu}{\Gamma(\nu+1)}\,,
}{eq:ap2f}
to approximate the heat kernel.  We find
\eq{
K_0(r,t,r,t;\si) \sim \si^{-\frac{d}{2}}e^{-r^2/2\si}
}{eq:ap2g}
giving the desired spectral dimension
\eq{
D_{\rm spec} = d \,.
}{eq:ap2h}

\begin{figure}
\centering
\epsfig{file=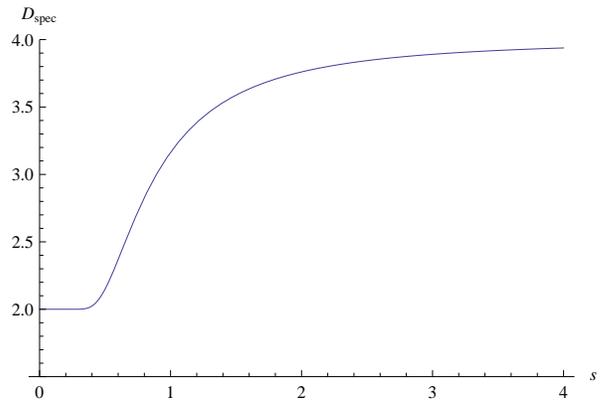,width=0.9\linewidth}
\caption{Spectral dimension $D_{\rm spec}$ for $d=4$ as function of $s$}
\label{fig:2}
\end{figure}

For $d=4$, the heat kernel \eqref{eq:ap2e} can be expressed in
terms of elementary functions, and the crossover between the ``small
$\si$'' and ``large $\si$'' regimes can be investigated analytically.
This crossover happens quite rapidly, at relatively small values: if
we define a dimensionless variable $s = \sqrt{\si}/r$, we find
that $D_{\rm spec}$ rises from very nearly two at $s=0.4$ to very
nearly four at $s=4$ (see Fig.~\ref{fig:2}).

Now let us consider the generalization to non-integer dimension.  
The dimension entered our derivation only in the behavior of the dilaton, 
$X\sim r^{d-2}$, and in the $r$ dependence of the potential term in 
\eqref{eq:ap2b}.  But nothing in the derivation required that $d$
be an integer.  Hence if the dilaton behaves as $X\sim r^{d-2}$ and 
the potential term has a lowest eigenvalue of zero, the spectral
dimension is $d$, whether $d$ is an integer or not.  Equivalently,
$D_{\rm spec} = 2 +  \extd\,(\ln X)/\!\extd\, (\ln r)$, 
just as suggested in \eqref{eq:ap2i}.
For a spectral dimension that varies slowly with scale --- more
precisely, one that varies slowly compared to the low-lying eigenfunctions
of \eqref{eq:ap2b} --- this should remain a good approximation.  As we
show in appendix \ref{app:A} below, this certainly seems to be the case 
in a simple, explicit model, in which the spectral dimension calculated by
\eqref{eq:ap2i} and the effective dimension in the dilaton potential
agree well over a very large range of values of $X$.

\section{Paper-and-pencil example}\label{app:A}

A simple dilaton gravity model that realizes the features discussed 
in the main text can be obtained by choosing
\eq{
D(X) = 4 - \frac{5c/2}{c+X}
}{eq:15}
for the effective dimension in \eqref{eq:9}.  By construction, we have 
$D_{\rm IR}=4$ and $D_{\rm UV}=3/2$, as in the EDT result
\eqref{eq:2}.  The action is then given by \eqref{eq:6} with 
potentials
\eq{
U(X) = -\frac{X-3c/2}{2X^2-cX/2}\qquad 
V(X)=-X^5 \,\frac{X-3c/2}{(X-c/4)^6} \, ,
}{eq:14}
and a convenient choice of integration constants yields 
\eq{
w(X)=\exp{\{-Q(X)\}} = 2X^3(X-c/4)^{-5/2}\,.
}{eq:14a}
The deformed Schwarzschild black hole is then described by the 
line element \eqref{eq:angelinajolie} with Killing norm
\eq{
\xi(X) = 1 - \frac{2M\,(X-c/4)^{5/2}}{X^3}\,.
}{eq:16}

The dilaton field evaluated at the horizon, $X_h$, can be expressed 
in terms of the black hole mass $M$ by solving $\xi(X_h)=0$ numerically.
Applying the general results of \cite{Gegenberg:1994pv,Grumiller:2007ju},
we obtain a Hawking temperature
\begin{align}
T_{\rm H} = \left.\frac{w'(X)}{4\pi}\right|_{X=X_h} 
&= \frac{1}{4\pi}\,\frac{X_h^3-3cX_h^2/2}{(X_h-c/4)^{7/2}}
\label{eq:17} \\
&= \frac{1}{8\pi M}\,\left( 1 + \frac{5c}{8M^2} 
   + {\cal O}(c^2/M^4)\right)
\nonumber
\end{align}
The Bekenstein--Hawking entropy is
\eq{
S_{\rm BH} = \frac{2\pi X_h}{G_2} = 4\pi M^2 \left(1- \frac{5c}{16M^2} 
+ {\cal O}(c^2/M^4)\right)\,, 
}{eq:18}
where we set $G_2=2$ in order to obtain the entropy in 4-dimensional 
Planck units where $G_N=1$.  That this is the correct value of the 
2-dimensional gravitational coupling constant $G_2$ can be seen,
e.g., from Eqs.~(5)-(8) in \cite{Grumiller:2007wb}.
The specific heat is then 
\eq{
C = 2\pi\,\frac{w'}{w''}\Big|_{X=X_h} \!\!\!\!\!
= 4\pi\,\frac{X_h(X_h-c/4)(X_h-3c/2)}{-X_h^2+3cX_h+3c^2/2}\,.
}{eq:19}

For large black holes, those with $M^2\gg c$, the standard 
Schwarzschild results are recovered.  Thus, black hole 
evaporation initially follows rather precisely the semiclassical 
approximation.  For smaller black holes, however, the thermodynamic 
properties deviate appreciably from the semiclassical results.
In particular, in the interval $3c/2<X_h<c(3+\sqrt{15})/2$,
the specific heat is positive.  
At $X_h=c(3+\sqrt{15})/2$ it has a pole, indicating a 
Hawking--Page-like phase transition.  In the limit $X_h\to 3c/2$, 
the black hole temperature and specific heat both drop to zero, 
in accordance with the third law.  Specific heat scales linearly with 
temperature for small $T$, as in a degenerate Fermi gas.  The 
Sommerfeld constant scales like $c^{3/2}$, $C/T|_{T\to 0}\sim c^{3/2}$.

From \eqref{eq:15}, we see that the endpoint of Hawking 
evaporation corresponds to an effective dimension of $D=3$, 
as expected from the general discussion.  The entropy of the final 
extremal black hole is $S=3\pi c/2$, and thus depends on microscopic 
details.
A singularity occurs at 
$X=c/4$, corresponding to the universal result 
$D=2$.  This example not only realizes the general features discussed 
in the body of the paper, but also provides the first concrete dilaton 
gravity model for an evaporating Schwarzschild black hole with 
bounded Hawking flux, and indeed recovers the end state predicted in 
\cite{Grumiller:2003hq}.

\begin{figure}[b]
\centering
\epsfig{file=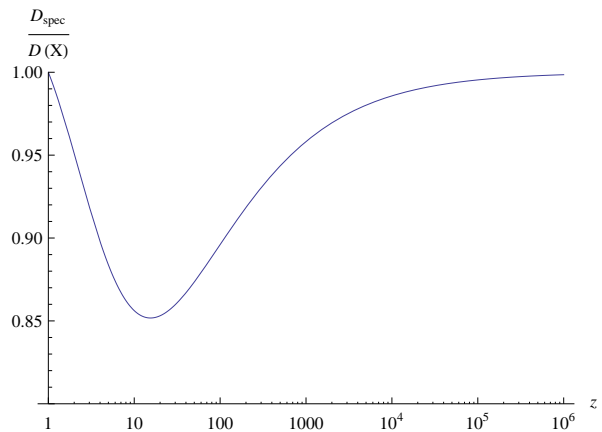,width=0.9\linewidth}
\caption{Ratio $D_{\rm spec}/D(X)$}
\label{fig:1}
\end{figure}

We can also compare the effective dimension \eqref{eq:15} to
the estimate of the spectral dimension in appendix \ref{app:B}.  There, 
we showed that if the spectral dimension varies slowly, it is 
approximately given by equation \eqref{eq:ap2i}.
From the solution for the dilaton \eqref{eq:lalapetz} with the
function $Q(X)$ as in \eqref{eq:14a}, we have
\eq{
r = \frac{\sqrt{c}}{16}\left[ 
   \sqrt{z-1}\left(8 + \frac{9}{z} - \frac{2}{z^2}\right) 
   - 15\arctan\sqrt{z-1} \right] \, ,
}{eq:x2}
where $z=\frac{4}{c}X$ is a rescaled dilaton field, and the integration
constant is chosen so that $r=0$ at the singularity.  Hence, again
using \eqref{eq:lalapetz}, we have
\eq{
D_{\rm spec} = \frac{1}{4}\frac{1}{(z-1)^2}\left[
    16z^2 - 7z + 6 - 15z^2\frac{\arctan\sqrt{z-1}}{\sqrt{z-1}}\right]\,.
}{eq:x3}

It is easy to check that $D_{\rm spec}$ approaches four for large $X$,
and that it is nonsingular at $z=1$ (i.e., $X=c/4$), approaching two.
Figure \ref{fig:1} shows the ratio $D_{\rm spec}/D(X)$ as a function 
of $z$ in a log-linear plot.  The ratio is nearly one for the entire range, 
with a maximum deviation of about 15\% around $z=15$, supporting 
our heuristic arguments in the body of this paper.


\end{document}